# Micro magnet location using spin waves

Michael Balinskiy and Alexander Khitun*

Electrical Engineering Department, University of California - Riverside, Riverside, CA, USA, 92521

**Abstract.** In this work, we present experimental data demonstrating the feasibility of magnetic object location using spin waves. The test structure includes a $Y_3Fe_2(FeO_4)_3$) (YIG) film with four micro-antennas placed on the edges. A constant in-plane bias magnetic field is provided by NdFeB permanent magnet. Two antennas are used for spin wave excitation while the other two are used for the inductive voltage measurement. There are nine selected places for the magnet on the film. The magnet was subsequently placed in all nine positions and spin wave transmission and reflection were measured. The obtained experimental data show the difference in the output signal amplitude depending on the magnet position. All nine locations can be identified by the frequency and the amplitude of the absolute minimum in the output power. All experiments are accomplished at room temperature. Potentially, spin waves can be utilized for remote magnetic bit read-out. The disadvantages and physical constraints of this approach are also discussed.

## I. Introduction

The ability to locate objects using transmitted or reflected waves is widely used in different technologies [1-3]. The position is related to the amplitude/frequency of the reflected/transmitted wave. It provides a convenient tool for non-contact object location. It may be possible to apply similar techniques to locate magnetic objects using spin waves. Spin wave is a collective oscillation of spins in a spin lattice around the direction of magnetization. Spin waves appear in magnetically ordered structures, and a quantum of spin wave is called a "magnon". The collective nature of spin wave phenomena manifests itself in relatively long coherence length, which may be order of the tens of micrometers in conducting ferromagnetic materials (e.g. $Ni_{81}Fe_{19}$) and exceed millimeters in non-conducting ferrites (e.g. YIG) at room temperature [4]. The latter makes it possible to build magnonic interferometers exploiting spin waves within the coherence length. For example, a Mach–Zehnder-type spin wave interferometer based on YIG structure was demonstrated in 2005 by M. Kostylev et al. [5]. The phase difference among the interfering spin waves was controlled by the magnetic field produced by an electric current flowing through a conducting wire under one of the arms. In general, spin wave dispersion depends on the strength and direction of the bias magnetic field [6]. Even a relatively weak (e.g. tens of Oersted) magnetic field produced by a micro-scale magnet placed on the top of magnonic waveguide may result in a prominent phase shift/amplitude change [7]. This phenomenon was utilized for magnonic holographic imaging of magnetic microstructures [8]. Here, we consider the feasibility of magnetic object location using spin waves. The rest of the paper is organized as follows. In the next Section II, we describe the experimental setup and present experimental data. The Discussion and Conclusions are given in Sections III and IV, respectively.

## II. Experimental data

The schematics of the experimental setup are shown in Fig.1. The cross-section of the device under study is shown in Fig.1(a). It consists of a permanent magnet made of NdFeB, a Printed Circuit Board (PCB) substrate with four short-circuited antennas, a ferrite film made of YIG, and a micro magnet that can be placed on different parts on top of the film. The permanent magnet is aimed to create a constant bias magnetic field. This bias magnetic field defines the frequency window as well as the type of spin waves that can propagate in the ferrite film. The strength of the field depends on the type of permanent magnet and the substrate thickness. The bias field is about 195 Oe and directed in-plane on the film surface. The photo of the PCB substrate with four antennas is shown in Fig.1(b). Each antenna is 6 mm long and 150 µm wide. The antennas are marked as 1,2,3, and 4 in the figure. The four antennas are placed on the side of a virtual square with a length of 11.4 mm. These antennas are aimed to excite and detect spin waves. The antennas are connected to the Programmable Network Analyzer (Keysight N5241A). The details of the spin-wave measurements with micro antennas can be found elsewhere [9,10].

The ferrite film is made of YIG. YIG was chosen due to the low spin wave damping. The thickness of the film is 31.5 µm. The saturation magnetization is close to 1750 G, the dissipation parameter ΔH = 0.6 Oe. The planar dimensions of YIG-film are significantly larger than a virtual square of antennas providing the total coverage of all microstrip antennas. The schematics in Fig.1(c) show the top view of the YIF film. It is shown a frame of reference, where one division corresponds to 0.75 mm. The red circles depict the nine possible positions of the magnet. Hereafter, the position of the magnet will be referred to this reference frame. The magnet has a disk shape whose diameter is about 1 mm and the thickness is 0.3 mm. The magnet is made of magnetic steel.

The first set of experiments is aimed to confirm spin-wave propagation through the film. These experiments are accomplished without a magnet on the top of the film. The collection of data showing S21 parameter measured with PNA is presented in Fig.2. In Fig.2(a), there are shown data for the case when the signal is excited by antenna #1 and detected by antenna #3 (see Fig.1(c)). The data are taken in the frequency range from 1.3 GHz to 2.9 GHz. The spectrum reveals Magnetostatic Surface Spin Wave (MSSW) propagation perpendicular to the direction of the bias magnetic field. In Fig.2(b), there is shown S21 parameter measured at PNA for the frequency range from 1.4 GHz to 2.2 GHz. The signal is excited by antenna #2 and detected by antenna #4. The spectrum reveals Backward Volume Magnetostatic Spin Wave (BVMSW) propagation directed along the bias magnetic field. In Fig.1(c), there are shown experimental data when two input antennas #1 and #3 operate simultaneously. The output signal is detected at antennas #2 and #4. The detected inductive voltage is a superposition of the two signals transmitted by MSSW and BVMSW. This configuration with two working input antennas is the most preferable for magnet location as it allows us to see the difference in spin wave propagation along the X and Y axes at a same time.

Next, the experiments with two operating input antennas (i.e., #1 and #2) and two output antennas (i.e., #3 and #4) were repeated for the different positions of the magnet. The magnet was sequentially placed in the nine positions marked with the red circles in Fig.1(c). The measurements were accomplished in the frequency range from 1.5 GHz to 3.0 GHz. In Fig.3, there are shown data obtained after the subtraction (i.e., signal without magnet) and filtering. The graph shows the change of the output transmitted power ΔP (i.e., measured by the two output antennas) as a function of frequency. There are nine curves of different colors that correspond to the nine magnet locations. The data reveal a prominent variation in the output power depending on the magnet position.

In Fig.4., there are shown data on spin wave signal transmission for the four selected cases corresponding to four selected positions of the magnet. The position of the magnet is shown in the inset. Fig.4(a) shows the normalized output power (i.e., after the subtraction) for the case when the magnet is placed in the center of the film. Figs. 4(b) and 4(c) show the output power for the magnet located on the corners of the film. In Fig.4(d), there are shown data for magnet shifted from the center towards the excitation antenna #1. The minimum of the signal transmission appears at different frequencies and reaches different amplitudes. That is the key result that allows us to conclude on the magnet position by the results of spin wave transport measurements. The summary of the experimental data obtained for different magnet locations is shown In Table I. The first two columns contain data on the magnet location while the last two columns show the frequency of the signal minimum and the normalized minimum amplitude. As one can see, there is a unique combination of frequency/minimum amplitude for each of the nine positions. The accuracy of the frequency measurements is 1 MHz. The accuracy of the output power measurements is 4.5 pW. All measurements are done at room temperature.

It is also interesting to investigate the change in the signal reflection (i.e., the S11 parameter) depending on the position of the magnet on the film. In Fig.5, there are shown data on signal reflection. The reflected signal is measured by antenna #1 and antenna #2. The change in the reflected power is shown after the subtraction (i.e., signal reflection without magnet) and filtering. The graph shows the change of the reflected power ΔP (i.e., measured by the two antennas) as a function of frequency. There are nine curves of different colors that correspond to the nine magnet locations. There is also a difference in the reflection where the minimum of the reflected power occurs at different frequencies and reaches different minimum values. There is a less difference in the frequency compared to the transmission signal. The difference in the amplitude of the reflected signal is prominent for the different locations of the magnet.

In Fig.6., there are shown data on spin wave signal reflection for the four selected cases corresponding to four selected positions of the magnet. The position of the magnet is shown in the inset. Fig.6(a) shows the normalized reflected power (i.e., after the subtraction) for the case when the magnet is placed in the center of the film. Figs. 4(b) and 4(c) show the reflected power for the magnet located on the corners of the film. In Fig.4(d), there are shown data for magnet shifted from the center towards the excitation antenna #1. The summary of the experimental data obtained for the reflected signal is shown In Table II. The first two columns contain data on the magnet location while the last two columns show the frequency of the signal minimum and the normalized minimum amplitude. There are two prominent minima in the reflected power that occur for several magnet positions (e.g., (2,0), (2,1), (1,2), and (2,2)). The minima in the reflected power appear on the same frequencies for the different magnet locations. Overall, the reflected spectra are less informative for the magnet location compared to the ones obtained for the transmitted signal.

### III. Discussion

There several observations we want to make based on the obtained experimental data. (i) The output power spectra for transmitted signal are not symmetric for symmetric position of the magnet. For instance, one can see in Table I that the minimum of the output power differs in frequency and amplitude for magnet placed in different corners. The movement of the magnet along the X or Y axes results in the different output power depending the direction of motion. This fact can be attributed to the asymmetry of the spin wave diffraction on the magnet and different dispersion of MSSW and BVMSW. The using of the same types of spin waves (e.g., only MSSW or only BVMSW) would smash the difference in the output characteristics. The calculation of the spin wave intensity profile over the film is a quite complicated computational task that goes beyond the scope of this work. The main focus of this work is the feasibility of magnet location via spin waves. (ii) The difference in the output power is quite prominent in the range of tens or a hundred of pW. It may be possible to recognize hundreds of magnet locations only by the amplitude of the output signal. The difference in the frequency of the output is also prominent, which provides an additional degree of freedom for magnet location. (iii) The experiments were accomplished on a relatively large template (e.g., the area of the film with four antennas is about 1 cm$^2$, the size of the magnet is about 1mm$^2$). These millimeter-scale dimensions are mainly defined by the wavelength of the spin waves. It is estimated that the wavelength of MSSW is about 0.5 mm. The wavelength of BVMSW is about 0.5 mm as well. These large dimensions are possible due to the long coherence length of spin waves in YIG. There is a lot of room for scaling down and increasing the number of possible magnet positions. The scaling down will require the reduction of the spin wave wavelength to micrometer range. In this work, the wavelength of the spin waves is mainly defined by the thickness of the YIG-film and the size of the micro-antennas. There should be a different mechanism for micrometer wavelength spin wave generation. For example, spin waves can be excited and detected by synthetic multiferroics [11]. However, this technique remains mainly unexplored.

The ability to search for a number of possible magnet positions is the most appealing property of the described approach to magnet location using spin waves. In contrast to the existing technologies based on magnetoresistance measurements, it does not require any physical contact between the magnet and the sensing element. Overall, it may provide a fundamental advantage over the existing practices for magnetic bit addressing and read-out. The spin wave location technique may be further extended by increasing the number of input/output ports [12] or/and exploiting spin wave interference [13]. It would be of great interest to validate the possibility to identify multiple magnet configurations (i.e., configuration of several magnets on selected locations). That would significantly enhance the read-out information capacity and lead to a new class of magnetic memory.

There are several physical limitations and constraints inherent in the spin wave approach. The recognition of the magnet position requires the scan over a frequency range to find the location of the absolute minimum. It complicates the whole search procedure and requires additional resources for input frequency modulation. The accuracy of output power measurements is another physical constraint that limits the number of possible magnet locations or magnets configuration to be recognized. The physical origin of the prominent change in the signal transmission/reflection depending on the position of the magnet is not well understood. There are multiple factors that affect spin wave propagation (e.g., non-uniformity of the bias magnetic field, non-uniformity of the magnetic field produced by the magnet, etc). The position of the magnet may also affect the generation of spin waves by the input antennas. One of the critical concerns is related to the scalability of the proposed approach. On the one hand, quite a large propagation length of spin waves (i.e., up to 1 cm) at room temperature in YIG serves as the base for further device scaling. On the other hand, it is not clear if the difference in the signal transmission will be still recognizable for nanometer scale magnets.

**IV. Conclusions**

We present experimental data showing the change of spin wave transmitted and reflected signal depending on the magnet position on the film. Overall, the data show a prominent variation in the frequency and the amplitude of the signal depending on the magnet position. It is possible to conclude on the location of the magnet (i.e., one of the nine pre-selected positions) based on the spin wave measurements. All experiments are accomplished at room temperature. It demonstrates the practical feasibility of using spin wave for magnetic object location. It may be utilized for magnetic bit addressing and read-out. The physical origin of the prominent signal modulation is not clear. There are multiple factors affecting spin wave propagation/generation/detection which need further investigation. The experiments are accomplished on a relatively large template with millimeter-sized antennas. The main practical challenge toward nanometer magnet location is associated with a short-wavelength spin wave generation and detection.

**Author Contributions**
M.B. carried out the experiments. A.K. conceived the idea of magnet location using spin wave and wrote the manuscript. All authors discussed the data and the results and contributed to the manuscript preparation.

**Competing financial interests**
The authors declare no competing financial interests.

**Data availability**
All data generated or analyzed during this study are included in this published article.

**Acknowledgment**
This work was supported by the National Science Foundation (NSF) under Award # 2006290.

**Figure Captions**

Figure 1. (a) Schematics of the test structure. It consists of a permanent magnet made of NdFeB, a Printed Circuit Board (PCB) substrate with four short-circuited antennas, a ferrite film made of YIG, and a micro magnet that can be placed on different parts on top of the film. (b) The photo of the PCB substrate with four antennas. Each antenna is 6 mm long and 150 μm wide. The antennas are marked as 1,2,3, and 4 in the figure. The antennas are connected to the Programmable Network Analyzer (Keysight N5241A. (c) The top view of the YIF film. It is shown a frame of reference, where one division corresponds to 0.75 mm. The red circles depict the nine possible positions of the magnet. The magnet has a discoidal shape whose diameter is about 1 mm and the thickness is 0.3 mm. The magnet is made of magnetic steel.

Figure 2. The collection of data showing S21 parameter measured with PNA without a magnet. (a) Experimental data for the case when the signal was is excited by antenna #1 and detected by antenna #3. The data are taken in the frequency range from 1.3 GHz to 2.4 GHz. (b) Data obtained for the case when the signal is excited by antenna #2 and detected by antenna #4. Data are collected in the frequency range from 1.4 GHz to 2.2 GHz. (c) Experimental data for the case when two input antennas #1 and #3 operate simultaneously. The output signal is detected at antennas #2 and #4.

Figure 3. Experimental data on spin wave transmission collected for nine positions of the magnet on the top of YIG film. There are nine curves of different colors that correspond to the nine magnet locations. The measurements were accomplished in the frequency range from 1.5 GHz to 3.0 GHz. The data are after the subtraction (i.e., signal without magnet) and filtering.

Figure 4. Experimental data on spin wave signal transmission for the four selected positions of the magnet. The position of the magnet is shown in the inset. (a) data for magnet placed in the center of the film; (b) data for magnet placed in the left top corner; (c) magnet is placed in the right down corner; (d) magnet is shifted from the center towards antenna #1.

Table I. Summary of the experimental data showing the frequency and the amplitude of the absolute minimum of the transmitted signal for nine magnet locations. The first two columns contain data on the magnet location while the last two columns show the frequency of the signal minimum and the normalized minimum amplitude.

Figure 5. Experimental data on spin wave reflection collected for nine positions of the magnet on the top of YIG film. There are nine curves of different colors that correspond to the nine magnet locations. The measurements were accomplished in the frequency range from 1.5 GHz to 3.0 GHz. The data are after the subtraction (i.e., signal without magnet) and filtering.

Figure 6. Experimental data on spin wave signal reflection for the four selected positions of the magnet. The position of the magnet is shown in the inset. (a) data for magnet placed in the center of the film; (b) data for magnet placed in the left top corner; (c) magnet is placed in the right down corner; (d) magnet is shifted from the center towards antenna #1.

Table II. Summary of the experimental data showing the frequency and the amplitude of the absolute minimum of the reflected signal for nine magnet locations. The first two columns contain data on the magnet location while the last two columns show the frequency of the signal minimum and the normalized minimum amplitude.

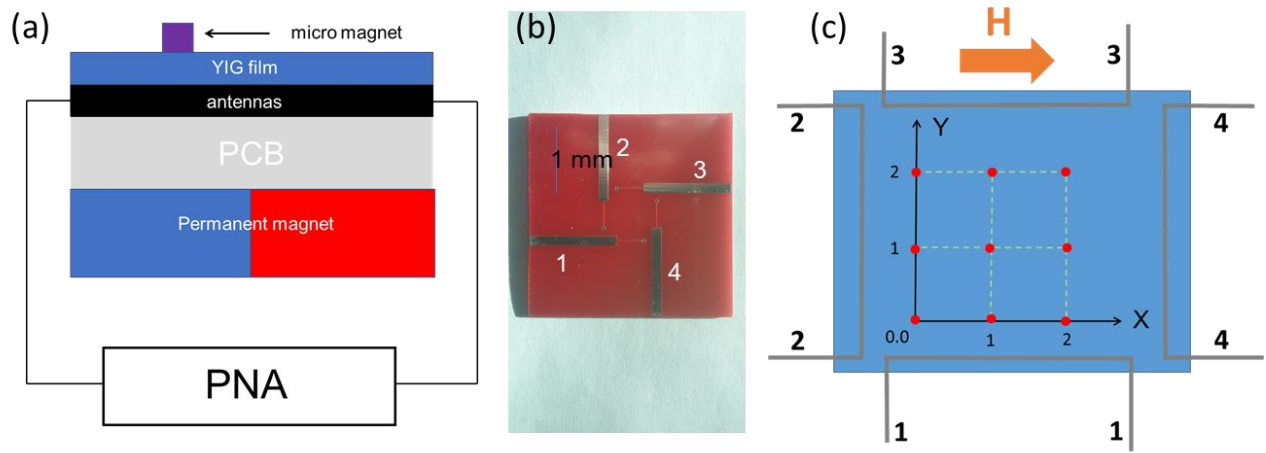

Figure 1

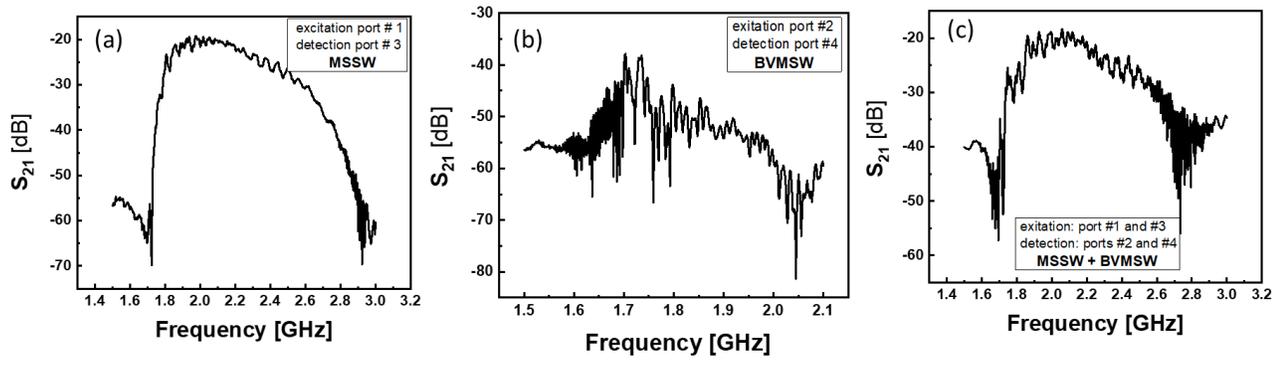

Figure 2

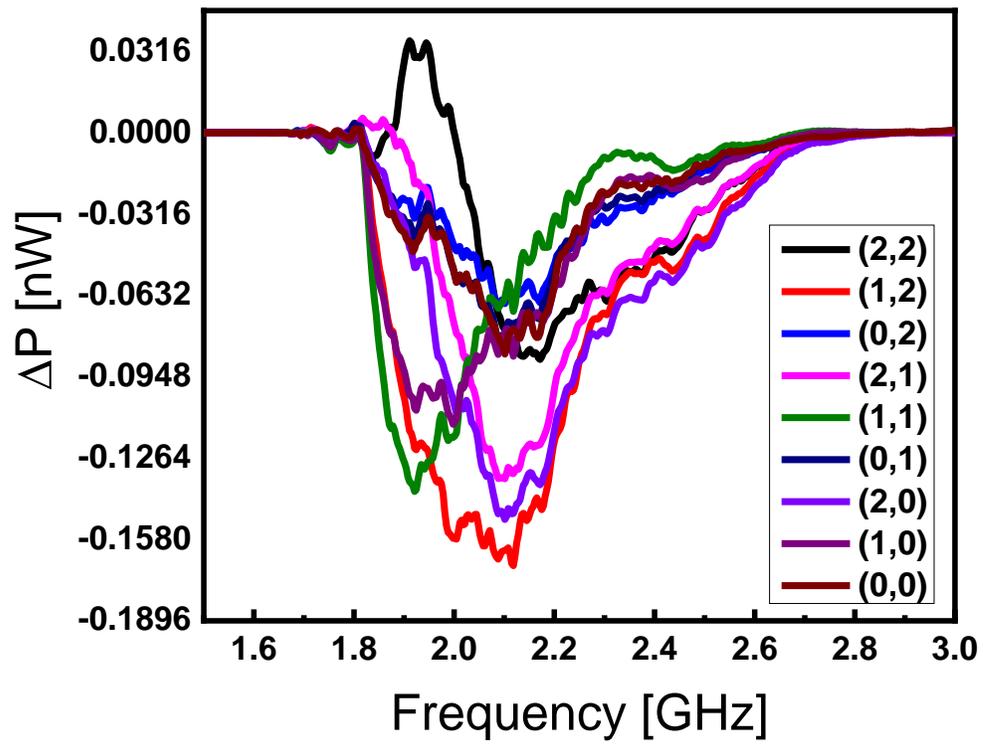

Figure 3

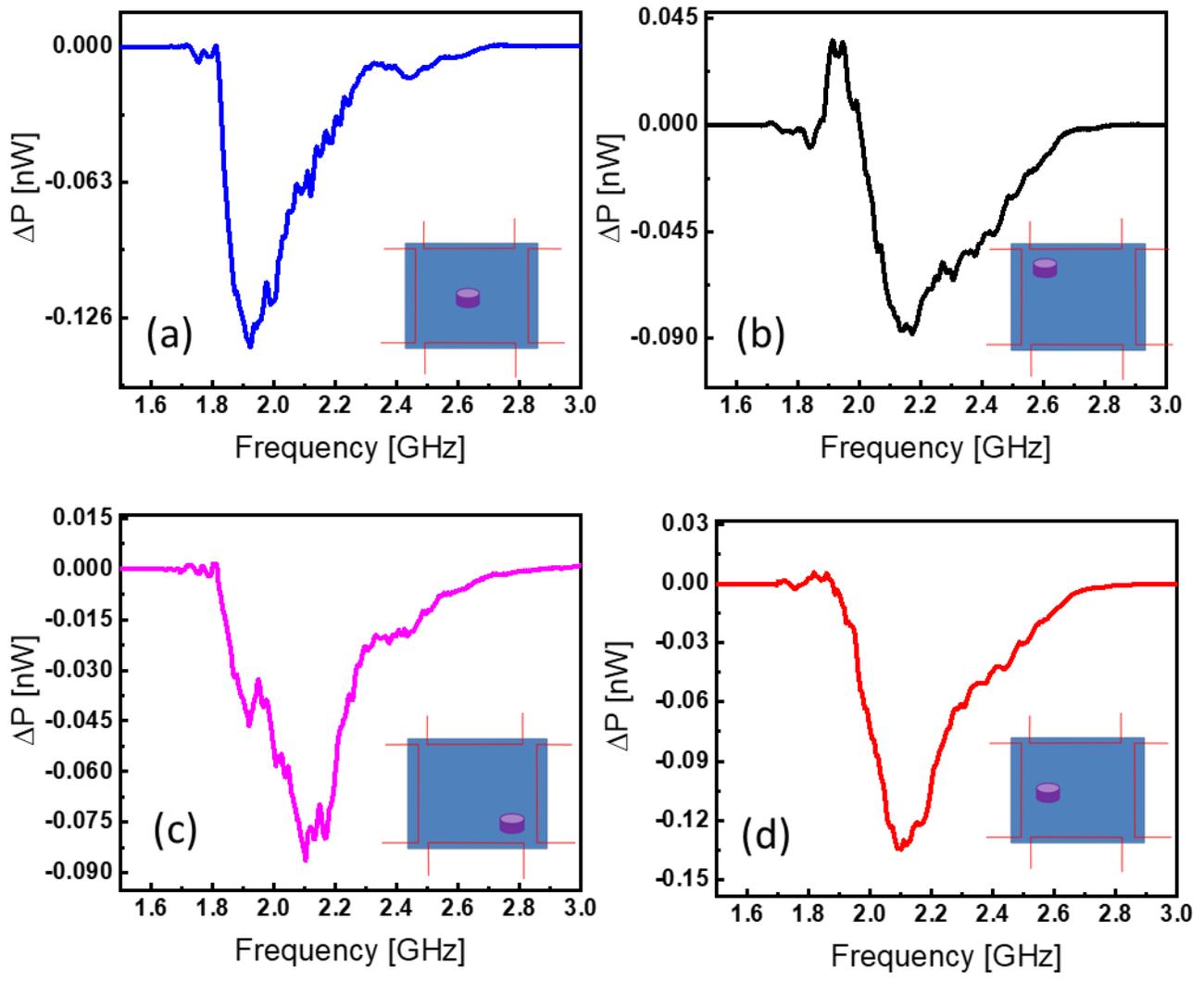

Figure 4

| X | Y | | ƒ [GHz] | ΔP [pW] |
|---|---|---|---|---|
| 0 | 0 | | 2.106 | -150 |
| 1 | 0 | | 1.996 | -113 |
| 2 | 0 | | 2.109 | -87 |
| 0 | 1 | | 2.099 | -134 |
| 1 | 1 | | 1.921 | -140 |
| 2 | 1 | | 2.099 | -79 |
| 0 | 2 | | 2.175 | -87 |
| 1 | 2 | | 2118 | -168 |
| 2 | 2 | | 2.091 | -66 |

Table I

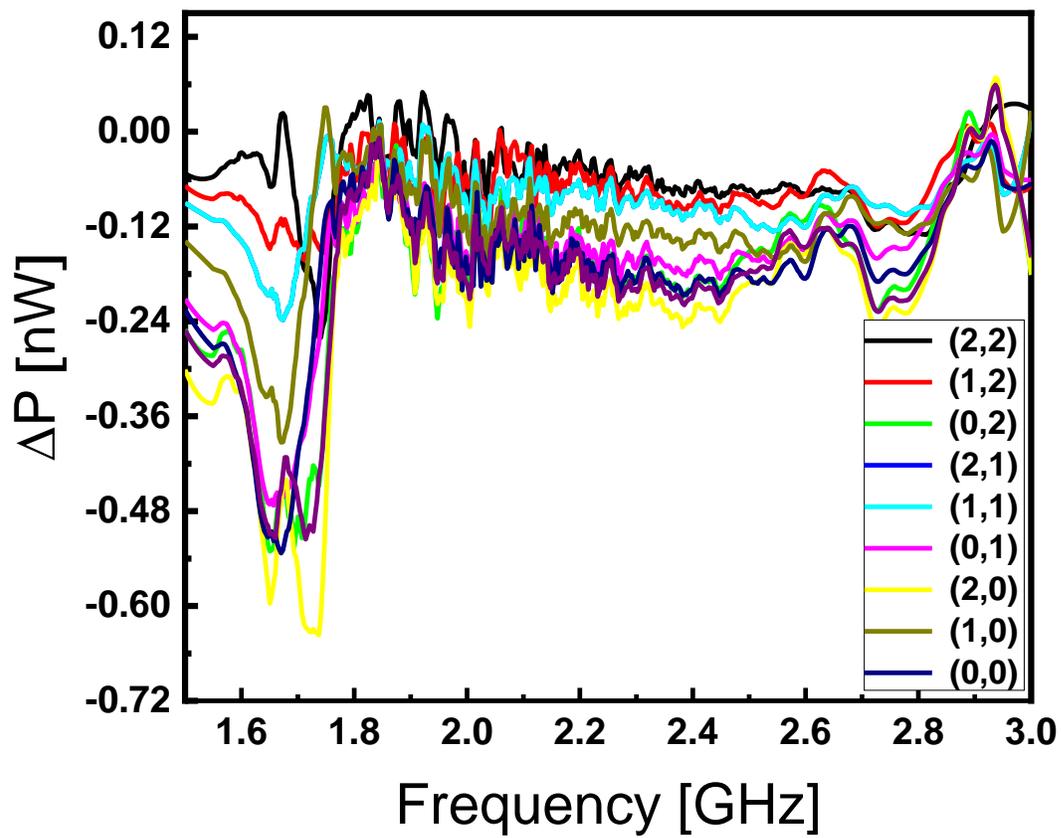

Figure 5

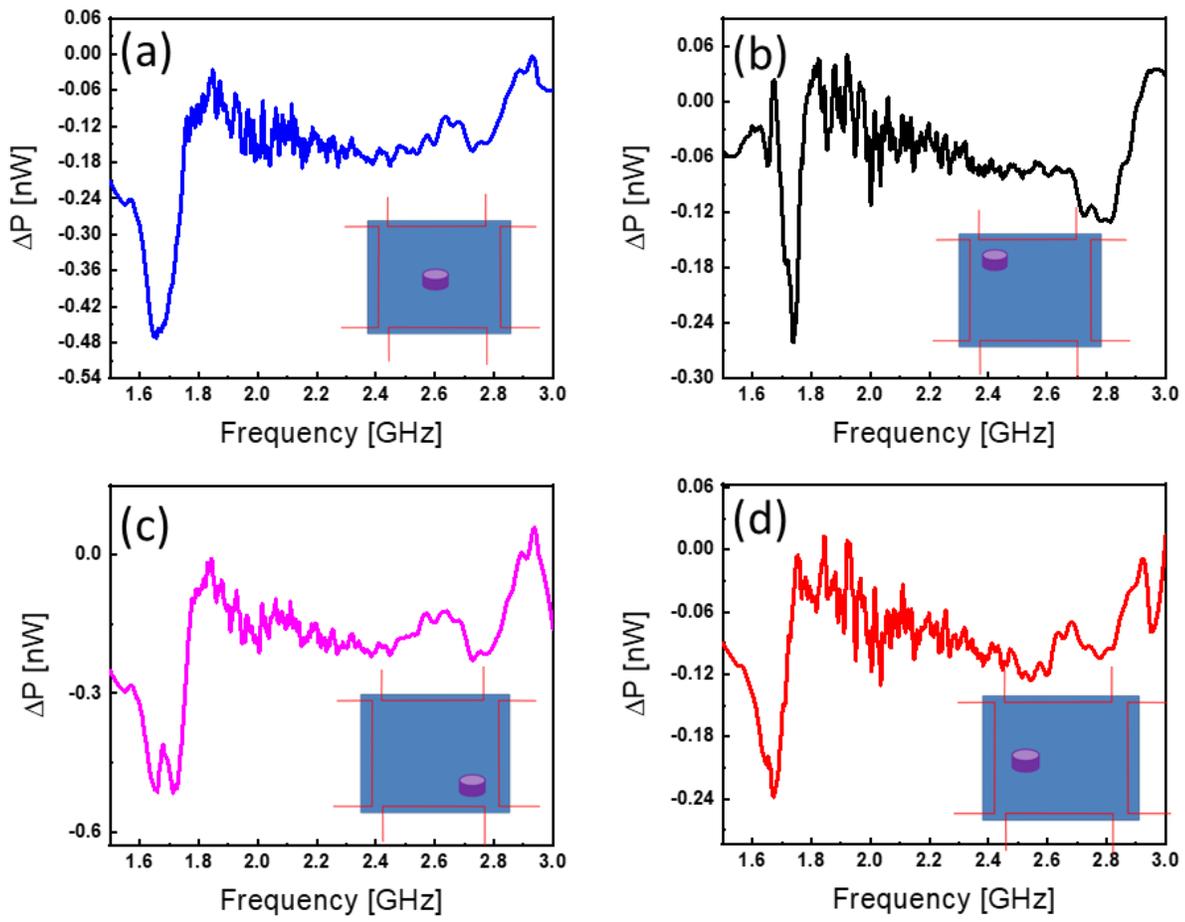

Figure 6

| X | Y | | ƒ [GHz] | | ΔP [pW] | |
|---|---|---|---------|-------|---------|------|
| 0 | 0 | | 1.674 | | -396 | |
| 1 | 0 | | 1.666 | | -532 | |
| 2 | 0 | | 1.659 | 1.716 | -515 | -516 |
| 0 | 1 | | 1.675 | | -239 | |
| 1 | 1 | | 1.652 | | -474 | |
| 2 | 1 | | 1.652 | 1.730 | -601 | -636 |
| 0 | 2 | | 1.737 | | -253 | |
| 1 | 2 | | 1.652 | 1.709 | -150 | -162 |
| 2 | 2 | | 1.652 | 1.702 | -531 | -525 |

Table II